\documentclass[12pt]{article}
\usepackage{latexsym, amssymb, amsmath}

\newtheorem{cor}{Corollary}[section]
\newtheorem{theo}{Theorem}[section]

\newtheorem{lem}{Lemma}[section]

\newenvironment{dem'}{\noindent \bf Proof: \rm}{$\quad \Box$}

\def \UN{\hbox{1\hskip -3.2pt I}}

\begin{document}
\title{Irregular sets and Central Limit Theorems for dependent triangular arrays}

\author{Beatriz Marr\'on $\, ^{*}$,  Ana Tablar $\, ^{1},$ \begin{footnote}
{Departamento de Matem\'atica. Universidad Nacional del Sur. $^1$ Corresponding author: actablar@uns.edu.ar}
\end{footnote}}\date{May 10, 2001}
\maketitle

{\bf Abstract}\\ In previous papers, we studied the asymptotic behaviour of $S_N(A,X)=(2N+1)^{-d/2}\sum_{n \in A_N} X_n,$ where $X$ is a centered, stationary and weakly dependent random field, and $A_N=A \cap [-N,N]^d$, $A \subset \mathbb{Z}^d$. This leads to the definition  of asymptotically measurable sets, which enjoy the property that $S_N(A;X)$ has a Gaussian weak limit for any $X$ belonging to a certain class. Here we extend this type of results to the case of weakly dependent triangular arrays and present an application of this technique to regression models. Indeed, we prove that CLT and related results hold for $X_n^N=\varphi(\xi_n^N,Y_n^N),\; n \in \mathbb{Z}^d$,  where $\varphi$ satisfies certain regularity conditions, $\xi$ and $Y$ are independent random fields, $\xi$ is weakly dependent and  $Y$ satisfies some Strong Law of Large Numbers.\\


\section{Introduction}

The notion of an  \lq\lq asymptotically measurable set\rq\rq$\,$ was introduced in \cite{Perera1} and \cite{Perera2}, and it was motivated by some statistical problems concerning random fields.\\
\indent Let us denote $\mathbb{Z}^d$ the lattice of points of $\mathbb{R}^d$ with integer coordinates. A subset $A$ of $\mathbb{Z}^d$ is said to be {\it asymptotically measurable} (AM), if for each $n \in \mathbb{Z}^d$, the limit, as $N$ tends to infinity, of $H_N(n;A)=\frac{card\{A_N\cap \left( n+A_N\right)\}}{(2N+1)^d}$  exists, where $A_N=A\cap [-N,N]^d$; furthermore, if we denote $H(n;A)$ this limit, it satisfies $0 < H(n;A) < 1$.

We denote $M(\mathbb{Z}^d)$ the class of asymptotically measurable sets. Sets with regular borders (in the sense that their borders are negligible), periodic sets and certain random sets are examples of elements of $M(\mathbb{Z}^d)$.

The class of centered, stationary, with finite second moment random fields which satisfy certain weak dependence conditions is denoted by $\mathbb{F}$, then $S_N(A,X)=\frac{1}{\sqrt{(2N+1)^d}}\sum_{n \in A_N} X_n$ has a non-trivial weak limit for any $X \in \mathbb{F}$ if and only if $A \in M(\mathbb{Z}^d)$, this is the main property of this class of sets.

For statistical purposes, a generalization of the notion of AM set is needed. We say that a collection $\{A^i:i=1,...,r\}$ of subsets of $\mathbb{Z}^d$ is an {\it asymptotically measurable collection} (AMC) if
$$ \lim_N\; H_N(n;A^i,A^j)= H(n;A^i,A^j)\; \forall n \in \mathbb{Z}^d,\;  i,j=1,...,r,$$
where
$ H_N(n;A^i,A^j)= \frac{card\{A_N^i \cap (n+A_N^j)\}}{(2N+1)^d}.$\\

Now consider $X=(X^1,...,X^r),$ a $\mathbb{R}^r$-valued, centered, stationary and weakly dependent random field, and define
$$M_N(A^1,...,A^r;X^1,...,X^r)= \left(S_N(A^1,X^1),...,S_N(A^r,X^r)\right),$$
then $M_N(A^1,...,A^r;X^1,...,X^r)$ converges weakly for any $X$ in a suitable class, if and only if $A^1,...,A^r$ is an AMC.\\

For instance, to be more precise, take $\|x\|=max_{1\leq i \leq r} |x^i|,\; x=(x^1,...,x^r)\; \in \mathbb{R}^r$, and let $X=(X_n)_{n\in \mathbb{Z}^d}\in \mathbb{F}$ be a random fields such that the following conditions hold:
\begin{itemize}
\item [{$(C1)$}] Let us call $r^X(k)=E\{X_0X_k\}$, then $\sum_{k \in \mathbb{Z}^d} |r^X(k)|<\infty.$
\item [{$(C2)$}] Let us define $X^J$, the truncation by $J$ of the random field $X$, that is
$X_n^J=X_n \UN_{\{\|X_n\|\leq J\}}- E\{X_n \UN_{\{\|X_n\|\leq J\}}\}.$
\begin{itemize}
\item [{$(i)$}] There exists no negative numbers $\rho(1), \rho(2), \cdots $ such that, for all $k \in \mathbb{Z}^d, J \geq 0$, $\sum_{k \in \mathbb{Z}^d}\rho(k)<\infty$, and  $|r^{X^{J}}(k)|<\rho(k).$
\item [{$(ii)$}] There exists a sequence $b(J)$ such that $\lim_{J \to \infty} b(J)=0$ and for each $A \subset \mathbb{Z}^d$, we have

$$ E\left\{(S_N(A,X)-S_N(A,X^J))^2\right\} \leq b(J) \frac{card(A_N)}{(2N+1)^d}.$$
\end{itemize}

\item[{$(C3)$}] For each $J>0,$ there exists a real number $C(X,J)$ such that for all $N \in \mathbb{N}, \; A \subset \mathbb{Z}^d$ we have $E\left\{(S_N(A,X^J))^4\right\}\leq C(X,J) \left(\frac{card(A_N)}{(2N+1)^d}\right)^2.$
\item [{$(C4)$}] There exists a bounded real function $g$ and a sequence $d(J)$ with $\lim_{J \to \infty} d(J)=0$ such that
$$ \left \vert E \left\{ \exp[it \, S_N(A\cup B,X^J)]\right\}- E\left\{ \exp[it \, S_N(A,X^J)]\right\}\cdot E\left\{\exp[it\, S_N(B,X^J)]\right\} \right \vert \hspace{2cm}$$
$$\hspace{10.5cm}\leq d(J)g(t), \; t \in \mathbb{R},$$
holds for any $A,B\subset \mathbb{Z}^d$ that satisfy $ dist\,(A,B)\geq J$.
\end{itemize}
\begin {theo}
If $A$ is an AM set and conditions (C1)-(C4) hold, then
$$S_N(A,X){\stackrel{w}{\longrightarrow}} N(0,\sigma^2(A,X)),$$
where $\sigma^2(A,X)=\sum_{n \in \mathbb{Z}^d}r^X(n)H(n;A).$
\end {theo}

The proof of this theorem is obtained by Bernshtein \lq\lq big and small blocks\rq \rq $\,$ method \cite{Bernshtein}, and it is similar to the proof of Proposition 2.2 in \cite{Perera3}.

Some final remarks on general notation we use all along this paper:

\begin{itemize}
\item The weak convergence of probability measures is denoted by \lq\lq${\stackrel{w}{\longrightarrow}}$\rq\rq.

\item The symbol \lq\lq 0\rq\rq represents both, the real zero and the zero element of $\mathbb{R}^d$; the context will make its meaning clear.

\item $N(\mu, \sigma^2)$ denotes a Gaussian distribution with mean $\mu$ and variance $\sigma^2$.

\item $card (A)$ is the cardinal of $A$.
\item $[x]$ is the integer part of the number $x$.
\item $X\approx Y$ means that $X$ and $Y$ has the same distribution.
\end{itemize}

\section{The central limit theorem for triangular arrays}

Here we shall deal with a special case of random fields, the triangular array. A triangular array is a double sequence of random variables $X_n^N$, $n \in \mathbb{Z}^d$, $N\in \mathbb{N}$, and the random variables in each row are independent. The  purpose of this paper is to establish a CLT for weakly dependent triangular arrays.

\begin{theo}
Let $\displaystyle X^N = \left(X_n^N \right)_{n \in \mathbb{Z}^d}$ be a triangular array that satisfies

\begin{itemize}

\item[$(H1)$] For all $\, N \in \mathbb{N}, \ X^N$ is a stationary, centered, with finite second moment random fields, that satisfies $\sum_{k \in \mathbb{Z}^d} \left| r^{X^N} \left( k \right) \right| < \infty, $ where $r^{X^N}(k)=E \left\{ X_0^N X_k^N \right\}.$

\item[$(H2)$] For all $\, J > 0 $ we define
$\displaystyle X_n^{N,J}= X_n^N \UN_{\{ \| X_n^N \| \leq J \}} - E \left\{ X_n^N \UN_{\left\{ \| X_n^N \| \leq J \right\}} \right\} $
and let us suppose that
\begin{itemize}

\item [$(i)$] There exists $\; \rho (k) \geq 0 $ such that $ \sum_{k \in \mathbb{Z} ^d} \rho(k) < \infty$ and for any $\, k \in \mathbb{Z} ^d $, $N \in \mathbb{N}$
$$\left| r^{X^{N,J}}(k) \right| \leq \rho(k).$$

\item [$(ii)$] There exists a sequence $\; b(J)$ such that $\lim_{J\to +\infty } b(J)=0$
and for any $\, B \subset \mathbb{Z}^d$, $N \in \mathbb{N}$
$$ E\left\{ {\left[ {S_N\left( {B,X^N} \right)-S_N\left( {B,X^{N,J}} \right)} \right]^2} \right\}\leq b\left( J \right)\, {{card\left( {B _N} \right)} \over {\left( {2N+1} \right)^d}}.$$

\end{itemize}

\item[$(H3)$] For all $\, J > 0 $,  there exists $C(J) < \infty$ such that for all  $B \subset \mathbb{Z}^d ,$ and for all $ N \in \mathbb{N}$
$$ E\left\{ {S_N\left( {B,X^{N,J}} \right)^4} \right\}\leq C\left( J \right)\, \left( {{{card\;B_N} \over {\left( {2N+1} \right)^d}}} \right)^2.$$

\item[$(H4)$] There exists a decreasing real function $\; h : \mathbb{R} ^+ \rightarrow \mathbb{R}^+$ such that \mbox{$\lim_{x \to + \infty} h(x)=0$} and a real function $g(J,t)$ such that for all fixed \mbox{$J>0$}, $g$ is bounded on the second variable,
$\sup_{t\in \mathbb{R} }g\left( {J,t} \right)=g_J<\infty $, such that
$$ \left \vert E \left\{ \exp[it \, S_N(A\cup B,X^{N,J})]\right\}- E\left\{ \exp[it\, S_N(A,X^{N,J})]\right\}\, E\left\{\exp[it\, S_N(B,X^{N,J})]\right\} \right \vert \hspace{2cm}$$
$$\hspace{10cm}\leq g\left( {J,t} \right)\,h\left(  dist \left( {A,B} \right) \right),$$

for any disjoint sets$\, A,B \subset \mathbb{Z}^d$, for all $ N \in \mathbb{N}, \; \, t \in \mathbb{R}.$

\item[$(H5)$] For all $\, k \in \mathbb{Z}^d, \; J>0 , \; there \; exists \; \gamma^J(k), \; \gamma(k)$ such that $$\lim \limits_{N \to + \infty} r^{X^{N,J}}(k) = \gamma^J(k)\qquad and \qquad \lim \limits_{J \to + \infty} \gamma ^J (k) = \gamma(k).$$

\end{itemize}
If $A$ is an AM set, then $S_N\left( {A,X^N} \right) \mathop\rightarrow\limits_N^\omega N\left( {0,\sigma ^2\left( A \right)} \right)$,
with $\;\displaystyle\sigma ^2\left( A \right)=\sum_{k\in \mathbb{Z}^d } {\gamma \left( k \right)}H\left( {k;A} \right).$
\end{theo}

\noindent \rm{The proof is based on the following steps:}

\begin{itemize}

\item[{}] \rm{First, constraint the problem to work with a bounded centered field, with the truncation proposed in (H2).}

\item[{}] \rm{Then, follow Bernshtein \lq\lq big and small blocks\rq\rq methods, so that the sum of variables over small blocks is negligible, and the sum of variables in two different large blocks is asymptotically independent.}
\end{itemize}

\begin{dem'}
We consider two nondecreasing sequences of positive integer $p_N$ and $q_N$ such that: $\lim_{N \to \infty}p_N = \lim_{N \to \infty} q_N = 0 \; \lim_{N \to \infty}\frac{q_N}{p_N} = 0, \; \lim_{N \to \infty}\frac{p_N}{N} = 0$ and
$ \lim_{N \to \infty}(k_N)^d .h\left( q_N \right)= 0$, where $k_N = \left [\frac{2N+1}{p_N+q_N}\right].$

We call $I_N(j) = [-N + j\left( p_N + q_N \right) ; -N + j \left( p_N + q_N \right) + p_N]$,$\;$ \mbox{$0 \leq j \leq k_N$}; $I_N = \bigcup\limits_{j=0}^{k_N} {I_N(j)}$ and $ \Delta_N = I_N^d$ is the union of $(k_N)^d$ disjoints $d$-cubes of side $p_N$
\begin{equation}
\Delta_N = \bigcup \limits_{\ell=1}^{(k_N)^d}\Delta_N(\ell),
\end{equation}
$card \; \left( \Delta_N (\ell) \right) = \left( p_N +1 \right)^d$, hence $card \; \left( \Delta_N \right) = (k_N)^d \left( p_N +1 \right)^d$. Even more, if $\ell \not=\ell', \; dist \left(\Delta_N(\ell), \Delta_N(\ell') \right) \geq q_N$.

We can decompose $S_N \left(A,X^{N,J} \right)$ as follows
\begin{equation}
S_N \left( A,X^{N,J} \right) = S_N \left( A \cap \Delta_N, X^{N,J} \right)+ S_N \left( A \cap \Delta_N^C, X^{N,J}\right).
\label{descop}
\end{equation}
As $X^{N,J}$ is a stationary process, by {\bf Lemma \ref{mom}} and $H2\, (i)$
\begin{eqnarray}
E \left\{ \left(S_N \left( A \cap \Delta_N^C , X^{N,J}\right) \right)^2\right\}
&\leq&
C \; \frac{card \left( \left( A \cap \Delta_N^C \right)_N \right)}{\left( 2N+1 \right)^d} \nonumber\\
&\leq& C \; \frac{card \left( \left( \Delta_N^C \right)_N \right)}{ \left(2N+1 \right)^d}\nonumber\\
&=& C \; \frac{\left( 2N+1 \right)^d - (k_N)^d \cdot \left( p_N+1 \right)^d}{\left( 2N+1\right)^d}\nonumber\\
&=& C \; \left[ 1 - \left( \frac{p_N+1}{p_N + q_N} \right)^d \right].
\label{L2}
\end{eqnarray}

By (\ref{L2}), the second term in (\ref{descop}) converges in $L^2$ to 0, therefore  it is enough to prove that $S_N \left( A \cap \Delta_N , X^{N,J} \right)$ converges weakly to a Gaussian law.

Let $Y_1^N, \ldots ,Y^N_{(k_N)^d}$ be a sequence of random independent variables such that
$Y_\ell^N \approx S_N \left( A \cap \Delta_N(\ell), X^{N,J} \right) \mbox{ for each } J >0.$ In order to show that $S_N \left( A \cap \Delta_N , X^{N,J}\right)\approx \sum_{\ell = 1}^{(k_N)^d}Y_\ell^N$, it is sufficient to prove

$$
\left\vert E \left\{\exp \left({it \; \sum \limits_{\ell = 1}^{(k_N)^d}S_N \left( A \cap \Delta_N (\ell), X^{N,J} \right)}\right) \right\} -  E\left\{ \exp \left({it \; \sum \limits_{\ell = 1}^{(k_N)^d} Y_\ell^N }\right)\right\}\right\vert \mathrel{\mathop{\longrightarrow}\limits_{N\to \infty }}   0.$$
Applying {\bf Lemma \ref{gonza}} with $Z_\ell=\exp\left({it \; S_N \left( A \cap \Delta_N (\ell), X^{N,J}\right)}\right)$ we have

\begin{flushleft}
$\lefteqn{\left\vert E \left\{\prod \limits_{\ell = 1}^{(k_N)^d}\exp \left({it \; S_N \left( A \cap \Delta_N (\ell), X^{N,J} \right)}\right) \right\} - \prod \limits_{\ell = 1}^{(k_N)^d}E\left\{\exp\left({it \; S_N \left( A \cap \Delta_N (\ell), X^{N,J} \right)}\right) \right\}\right\vert}$
\end{flushleft}

\begin{eqnarray}
\lefteqn{\leq \sum_{j=1}^{(k_N)^d-1} \left\vert E \left\{\prod_{\ell=j}^{(k_N)^d} \exp\left({it \; S_N \left( A \cap \Delta_N (\ell), X^{N,J} \right)}\right)\right\}\right.}\hspace{1cm}\nonumber \\
& &-\left.E\left\{\exp \left({it \; S_N \left( A \cap \Delta_N (j), X^{N,J} \right)}\right)\right\}\;E \left\{\prod_{\ell=j+1}^{(k_N)^d} \exp \left({it \; S_N \left( A \cap \Delta_N (\ell), X^{N,J} \right)}\right)\right\}\right\vert \nonumber \\
&\leq&\sum_{j=1}^{(k_N)^d-1}{g(J,t)\;h\left( dist\, \left(A \cap \Delta_N(j), \bigcup\limits_{\ell=j+1}^{(k_N)^d}A \cap \Delta_N(\ell)\right)\right)}\nonumber\\
&\leq&(k_N)^d \;g_J\; h(q_N) \longrightarrow 0 \;\; {\rm as}\;\; N \longrightarrow \infty,
\end{eqnarray}
by (H4) and as the distance between $ \Delta_N (\ell)$ is larger than $q_N$, hence $S_N \left( A \cap \Delta_N , X^{N,J}\right)$ has the same asymptotic distribution of $\sum_{\ell = 1}^{(k_N)^d}Y_\ell^N$. These random variables are a tringular array of independents copies of $S_N \left( A \cap \Delta_N (\ell),X^{N,J} \right)$, centered and with finite variance  $\sigma_{N}^2=E \left\{ Y_\ell^N \right\}^2=E \left\{ S_N \left( A \cap \Delta_N (\ell),X^{N,J} \right) \right\}^2$. By Lyapunov's central limit theorem if, $\frac{\sum \limits_{\ell=1}^{(k_N)^d}E \left[ \left(Y_\ell^N \right)^{2 + \delta} \right]}{\sigma^{2+\delta}}\mathop \rightarrow \limits_{N\to \infty } 0$  for some positive $\delta$, then
\begin {equation}
\sum \limits_{\ell = 1}^{(k_N)^d}Y_\ell^N  \mathop\rightarrow\limits_N^\omega N (0,\sigma_N^2).
\label {normalY}
\end {equation}
From $(H3)$
\begin{eqnarray*}
\sum \limits_{\ell=1}^{(k_N)^d} E \left\{ S_N \left( A \cap \Delta_N (\ell),X^{N,J} \right) \right\}^4
&\leq& \sum \limits_{\ell=1}^{(k_N)^d} C(J)  \left[ \frac{card \left(( A \cap \Delta_N(\ell)) _N\right)}{(2N+1)^d}  \right]^2 \\
&\leq& \sum \limits_{\ell=1}^{(k_N)^d} C(J)  \left[ \frac{card \left((\Delta_N(\ell)\right)_N) }{(2N+1)^d}  \right]^2\\
&=& C(J)  (k_N)^d \left( \frac{p_N+1}{2N+1} \right)^{2d}\\
&=& C(J)  \left[ \frac{(p_N+1)^2}{\left(p_N + q_N \right) \left(2N +1 \right)} \right]^d .
\end{eqnarray*}
The last equation tends to $0$ as $N\to \infty$, so the Lyapunov's condition  holds, with $\delta=2$, so (\ref{normalY}) follows.

By {\bf Lemma \ref{mom}} we can compute $\sigma_{N}^2$ as
\begin{eqnarray}
\sum \limits_{\ell=1}^{(k_N)^d} E \left\{ \left( S_N \left( A \cap \Delta_N(\ell), X^{N,J} \right) \right)^2 \right\}
&=& \sum \limits_{\ell=1}^{(k_N)^d} \sum \limits_{k \in \mathbb{Z}^d} r^{X^{N,J}}(k) H_N \left( k;A\cap \Delta_N(\ell) \right)\nonumber \\
&=& \sum \limits_{k\in \mathbb{Z}^d} r^{X^{N,J}}(k) \cdot \sum \limits_{\ell=1}^{(k_N)^d} H_N \left(k;A \cap \Delta_N(\ell) \right)\nonumber.\\
& &
\label{sigmajota}
\end{eqnarray}
By the other hand
$$H_N \left( k;A\cap \Delta_N(\ell) \right) \leq \frac{card \left( A \cap \Delta_N(\ell) \right)}{(2N+1)^d} \leq \frac{card  (\Delta_N(\ell)) }{(2N+1)^d}, $$
and
$$ 0 \leq \sum \limits_{\ell=1}^{(k_N)^d} H_N \left(k;A \cap \Delta_N(\ell) \right) \leq \sum \limits_{\ell=1}^{(k_N)^d} \frac{card (\Delta_N ( \ell))}{(2N+1)^d} \leq 1.$$
For each set $A$ we can decompose $A_N$ like $A_N=(A \cap \Delta_N) \cup (A \cap \Delta_N^C)$. For short, set \mbox{$D_N= A \cap \Delta_N$} and $C_N=A \cap \Delta_N^C$, then
\begin{eqnarray}
H_N \left(k;A\right)
&=& \frac{card \left( A_N \cap (k + A_N) \right)}{(2N+1)^d}\nonumber \\
&=& \frac{card \left( D_N \cap (k + D_N) \right)}{(2N+1)^d}+\frac{card \left( D_N \cap (k + C_N) \right)}{(2N+1)^d}+\frac{card \left( C_N \cap (k + C_N) \right)}{(2N+1)^d}\nonumber \\
& &+\frac{card \left( C_N \cap (k + D_N) \right)}{(2N+1)^d}.
\end{eqnarray}
As the last three terms above are bounded by $card \left(\Delta_N(\ell)\right)$,
\begin{equation}
H_N(k;A) \leq \frac{card \left( D_N \cap (k + D_N) \right)}{(2N+1)^d} + 3 \frac{card(\Delta_N^C)}{(2N+1)^d}.
\label{HN3}
\end{equation}
The second term in (\ref{HN3}) converges to $0$ if $N \to \infty$, then asymptotically $H_N(k;A) \simeq \frac{card \left(D_N \cap (k + D_N)  \right)}{(2N+1)^d}$, and
\begin{eqnarray}
\frac{card \left( D_N \cap  (k + D_N) \right)}{(2N+1)^d}
&=& \sum \limits_{\ell, \ell'=1}^{(k_N)^d} \frac{card \left\{ \left(A \cap \Delta_N (\ell) \right) \cap \left( k + (A \cap \Delta_N (\ell')) \right) \right\}}{(2N+1)^d}\nonumber\cr
&=& \sum \limits_{\ell=1}^{(k_N)^d} \frac{card \left\{ \left(A \cap \Delta_N(\ell) \right) \cap \left( k + ( A \cap \Delta_N(\ell)) \right) \right\}}{(2N+1)^d}\nonumber\cr
& & +\sum \limits_{\ell \not= \ell'} \frac{card \left\{ \left(A \cap \Delta_N(\ell) \right) \cap \left( k + ( A \cap \Delta_N(\ell')) \right) \right\}}{(2N+1)^d}.
\label{cardd}
\end{eqnarray}
For $N$ large enough such that $q_N > \left\| k \right\|,$ the second term in (\ref{cardd}) is \mbox{equal to $0$}, then
\begin{equation}
H_N(k;A) \simeq \sum \limits_{\ell=1}^{(k_N)^d} H_N(k;A\cap \Delta_N(\ell)).
\end{equation}
Since $A$ is measurable, $H_N(k;A) \mathrel{\mathop{\longrightarrow}\limits_{N\to \infty }} H(k;A),$ this holds
\begin{equation}
\sum \limits_{\ell=1}^{(k_N)^d} H_N \left(k;A \cap \Delta_N (\ell) \right) \mathrel{\mathop{\longrightarrow}\limits_{N\to \infty }} H(k;A).
\label{HN}
\end{equation}
Applying $(H5)$ and (\ref{HN}) in (\ref{sigmajota}), we have
$\sigma_N^2 \mathrel{\mathop{\longrightarrow}\limits_{N\to \infty }} \sigma_J^2(A)= \sum \limits_{k \in \mathbb{Z}^d} \gamma ^J(k) H(k;A).$
In summary we proved that
\begin{equation}
S_N \left( A, X^{N,J}\right)\mathop\rightarrow\limits_N^\omega N\left( 0, \sigma^2_J(A) \right).
\end{equation}

From hypotesis $(H2)\, i)$, \mbox{$\left|r^{X^{N,J}}\left( k \right) \right| \leq \rho \left( k \right), $} for any $k, N, J$, therefore, $\lim_{N \to \infty}\left|r^{X^{N,J}}\left(k \right) \right| \leq \rho \left( k \right)$.  So, by  $(H5)$ for any $ k, J$

$$\left|\gamma^J (k) \right| \leq \rho(k).$$
As $\sum_{k \in \mathbb{Z}^d}\rho(k) < \infty $
and since $0 \le H(k,A) \le 1$, applying the theorem of Dominated Convergence, we get that $ \sigma^2_J(A)$ is finite. Hence,
\begin {equation}
\lim_{J \to + \infty} \sigma^2_J \left( A \right) = \sigma^2 \left(A \right).
\label{sigma}
\end {equation}

For arbitraty $\epsilon >0$, by Tchebyshev inequality and $(H2) \,ii)$
\begin{eqnarray*}
P \left( \left| S_N  \left( A,X^{N,J} \right) - S_N \left( A,X^N \right) \right| > \epsilon \right)
&\leq& \frac{E \left\{ S_N \left(A,X^{N,J} \right) - S_N \left( A, X^N \right) \right\}^2}{\epsilon^2}\\
&\leq& \frac{b\left( J \right) card\left( A_N\right)}{(2N+1)^d\epsilon^2}\\
&\leq& \frac{b\left( J \right)}{\epsilon^2},
\end{eqnarray*}
then
$$\lim_{J \to + \infty} \limsup_{N \to \infty} \; P \left( \left|S_N \left(A,X^{N,J} \right)-S_N \left( A,X^N \right) \right| \geq \epsilon \right)=0,$$
and the theorem is proved.
\end{dem'}

\begin{cor}
Let $X^N = \left( X_n^N\right)_{n \in \mathbb{Z}^d}$ be a stationary, centered and $m$-dependent triangular array that satisfies the following conditions:

\begin{itemize}
\item[\it C1)] For each $N \in \mathbb{N},$ $E \left(X_0^N \right)^4 < {\cal C}$, ${\cal C} > 0$, a constant independent of $N$.

\item[\it C2)] The covariance function is uniformely bounded, that is
$$\left| r^{X^N} (k)  \right| \leq \rho (k) \hspace{1cm} 0 \leq \|k\| \leq m, \hspace{0.2cm} \forall \, N \in \mathbb{N}. $$

\item[\it C3)] For every $k \in \mathbb{Z}^d$, $J>0$, there exists $\; \gamma^J(k) < \infty$ and $\gamma (k) < \infty$ such that
$$\lim \limits_{N \to + \infty} r^{X^{N,J}}(k) = \gamma^J(k)\qquad and \qquad \lim \limits_{J \to + \infty} \gamma ^J (k) = \gamma(k).$$
\end{itemize}

If $A$ is AM, then $S_N\left( {A,X^N} \right) \mathop\rightarrow\limits_N^\omega N\left( {0,\sigma ^2\left( A \right)} \right)$,
with $\;\sigma^2(A)=\sum \limits_{\begin{array}{c} k \in \mathbb{Z}^d \\\|k\|\leq m \end{array}} \gamma(k) H(k;A).$
\end{cor}

\begin{dem'}
We set $X_n^{N,J}$ as in the theorem above, according to this theorem it is enough to show that conditions $(H1)$, $(H2)$, $(H3)$, and $(H4)$ hold.

The hypothesis $(H1)$ and $(H2) \;i)$ are direct from the fact that  \mbox{$ \left| r^{X^N} (k)\right| = 0$} if $\|k\| \leq m$.

Let us show $(H2)\; ii)$

$$E \left\{ \left[ S_N \left( B, X^N \right) - S_N \left( B , X^{N,J} \right) \right]^2 \right\} = E \left\{ \left[ \sum \limits_{n \in B_N} \frac{X_n^N - X_n^{N,J}}{\sqrt{\left(2N + 1 \right)^d}}\right]^2 \right\}. $$
Let us call $Y_n^{N,J}= \left( X_n^N - X_n^{N,J} \right)$, so
\mbox{$\sum \limits_{k \in \mathbb{Z}^d} \left| r^{Y^{N,J}}(k)\right| = \sum\limits_{\footnotesize{\begin{array}{c} k \in \mathbb{Z}^d \\ \|k\| \leq m \end{array}}} \left| r^{Y^{N,J}}(k) \right| < \infty$}.

Applying { \bf Lemma \ref{mom} }

\begin {eqnarray*}E \left\{ \left[ S_N \left( B , X^N \right) - S_N \left(B, X^{N,J} \right) \right]^2 \right\}
& =& E \left\{ S_N \left( B, Y^{N,J} \right)^2 \right\} \\
&\leq& \frac{card \left( B_N \right)}{\left( 2N + 1 \right)^d} \sum \limits_{\begin{array}{c} k \in \mathbb{Z}^d \\ \|k\| \leq m \end{array}} \left| r^{Y^{N,J}}(k) \right|,
\end {eqnarray*}
taking $b(J) = \sum \limits_{\begin{array}{c} k \in \mathbb{Z}^d \\ ||k|| \leq m \end{array}} \left| r^{Y^{N,J}} \right|$ to prove the hypothesis $(H2)\; ii)$ it is enough to show that $\left|r^{Y^{N,J}}\right| \longrightarrow 0 $ as $J \to \infty$, $\forall \, N \in \mathbb{N}$.

\begin{eqnarray*}
\left| r^{Y^{N,J}}(k) \right|
&=&\left| E \left\{Y_0^{N,J},Y_0^{N,J}\right\}\right|\\
&=&\left| {E\left\{ {X_0^N X_k^N} \right\}} -E\left\{ {X_0^N X_k^N \UN_{\left\{ {\left\| {X_k^N} \right\|\le J} \right\}}} \right\}\right.\\
& &\left.-E\left\{ {X_0^N \UN_{\left\{ {\left\| {X_0^N} \right\|\le J} \right\}}X_k^N} \right\} +  r^{X^{N,J}}(k)\right|.
\end{eqnarray*}
As applying Cauchy-Schwartz inequality, we have $E \left|X_0^N X_k^N \right|<\infty$, then by Dominate Convergence Theorem

$$ E \left\{ X_0^N X_k^N \UN_{\left\{\left\|X_k^N \right\| \leq J \right\}} \right\} \mathop{\longrightarrow} \limits_{J \to \infty} E \left\{ X_0^N X_k^N\right\},$$
$$ E \left\{ X_0^N \UN_{\left\{\left\|X_0^N \right\| \leq J \right\}} X_k^N\right\} \mathop{\longrightarrow} \limits_{J \to \infty} E \left\{ X_0^N X_k^N\right\},$$
and $r^{X^{N,J}}(k) \mathop {\longrightarrow} \limits_{J \to \infty} r^{X^N}(k)$, so
$b(J) = \sum \limits_{\begin{array}{c} k \in \mathbb{Z}^d \\ \|k\| \leq m \end{array}} \left| r^{Y^{N,J}}(k) \right| \mathop {\longrightarrow} \limits_{J \to \infty}0, \forall N \in \mathbb{N}.$

To show $(H3)$, let $B \subseteq \mathbb{Z}^d$ be an arbitrary set,
\begin{flushright}
\begin{eqnarray*}
\lefteqn{E \left\{ S_N \left( B, X^{N,J} \right)^4 \right\} = \frac{1}{(2N+1)^{2d}} \left[ \sum \limits_{i \in B_N} \tiny{E \left( X_i^{N,J} \right)^4} + 4 \sum \limits_{\scriptsize
{\begin{array}{c} i,j \in B_N \\ i \not= j \end{array}}} \tiny{E \left\{ \left( X_i^{N,J} \right)^3 X_j^{N,J} \right\}}\right.}\\
& &  + 6 \sum \limits_{\scriptsize
{\begin{array}{c} i,j \in B_N \\ i \not= j \end{array}}} \tiny{E \left\{ \left(X_i^{N,J} \right)^2 \left( X_j^{N,J} \right)^2 \right\}} + 12 \sum \limits_{\scriptsize
{\begin{array}{c} i,j,k \in B_N \\ i \not= j \not= k \end{array}}} \tiny{E \left\{ \left(X_i^{N,J} \right)^2 X_j^{N,J} X_k^{N,J} \right\}} \\
& &\left. + 24 \sum \limits_{\scriptsize
{\begin{array}{c} i,j,k,l \in B_N \\ i \not= j \not= k \not= l \end{array}}}\tiny{ E \left\{ X_i^{N,J} X_j^{N,J} X_k^{N,J} X_l^{N,J}  \right\}} \right].
\end{eqnarray*}
\end{flushright}
As $\left( X_i^{N,J} \right)^4 \leq \left( X_i^N \right)^4$, from $C1)$ we get

$$\sum \limits_{i \in B_N} E \left( X_i^{N,J} \right)^4 \leq card \left( B_N \right){\cal C}.$$

Applying Cauchy-Schwartz inequality and by the $m$-dependence, we get boundes for each term

\begin{eqnarray*}
\sum \limits_{ \scriptsize
{\begin{array}{c} i,j \in B_N \\ i \not= j \end{array}}} \small{ E \left\{ \left( X_i^{N,J} \right)^3 X_j^{N,J} \right\} }
&\leq & card \left( B_N \right) \left( card \left( B_N \right)-1 \right){\cal C}, \\
\end{eqnarray*}
\begin{eqnarray*}
 \sum \limits_{ \scriptsize
{\begin{array}{c} i,j \in B_N \\ i \not= j  \end{array}}}  E \left\{ \left( X_i^{N,J} \right)^2 \left( X_j^{N,J} \right)^2 \right\} &\leq&  card \left( B_N  \right) \frac{\left( card \left( B_N \right) -1 \right)}{2} {\cal C},
\end{eqnarray*}

$$\sum \limits_{ \scriptsize
{\begin{array}{c} i,j,k \in B_N \\ i \not= j \not= k \end{array}}}  E \left\{ \left( X_i^{N,J} \right)^2 X_j^{N,J} X_k^{N,J} \right\} \leq  \left( card \left( B_N \right) \right)^2 2 (2m+1)^d {\cal C},
$$

\begin{eqnarray*}
 \sum \limits_{\scriptsize
{ \begin{array}{c} i,j,k,l \in B_N \\ i \not= j \not= k \not= l \end{array}}}  E \left\{ X_i^{N,J} X_j^{N,J} X_k^{N,J} X_l^{N,J} \right\}
&\leq& \left( card \left( B_N \right) \right)^2 3 (2m+1)^d {\cal C}.
\end{eqnarray*}

We conclude that

$$E \left\{ S_N \left( B, X^{N,J} \right)^4 \right\}  \leq {\cal C}^\star \left( \frac{card \left(B_N \right) }{(2N+1)^d}\right)^2 \hspace{1cm} \forall J > 0.$$

To prove $(H4)$ let us note that the characteristic function of $S_N(A, X^{N,J})$ and the characteristic function of $S_N(B, X^{N,J})$ are independent random variables if $\,(A,B)>m$, then it is enough to consider $h(x)=0$ if $x>m$.
\end{dem'}

\section {Appendix}

\begin{lem}\label{mom}

 If $X=(X_n)_{n \in \mathbb{Z}^d}$ is a weakly stationary random field such that
for all  $n,$ $E(X_n)=0 \; , \; E(X_n^2) < \infty ,$
 then for any $B \subset \mathbb{Z}^d$,
$$E\left\{ {S_N\left( {B,X} \right)^2} \right\}=\sum\limits_{k\in \mathbb{Z}^d} {r^X\left( k \right)\cdot H_N\left( {k;B} \right)},$$
 with $ H_N(k;B)=\frac {card\; \left(B_N\cap(k+B_N)\right)}{(2N+1)^d}.$

 In particular, if $\sum_{k \in \mathbb{Z}^d} \vert r^X (k)\vert < \infty$ then $E\left\{ {S_N\left( {B,X} \right)^2} \right\}\leq C \,\frac {card\;\left( {B_N} \right)} {\left( {2N+1} \right)^d}.$
\end{lem}

\begin{lem}\label{gonza}
Let $ Z_1, Z_2, \cdots ,Z_n $ be a sequence of complex-valued  random variables such that $|Z_i|\leq 1 $ , for  all  $i $, then
$$\left\vert E \left\{\prod_{i=1}^{n} Z_i\right\}- \prod_{i=1}^{n} E \left\{Z_i \right\}  \right\vert\leq \sum_{j=1}^{n-1}\left\vert E \left\{\prod_{i=j}^{n} Z_i\right\}-E\{Z_j\}\;E \left\{\prod_{i=j+1}^{n} Z_i\right\}\right \vert.$$
\end{lem}

\begin{dem'}

$$\left\vert E \left\{\prod_{i=1}^{n} Z_i\right\}- \prod_{i=1}^{n} E \left\{Z_i \right\}  \right\vert\hspace{8cm} $$
\begin{flushright}
\begin{eqnarray*}
&\leq& \displaystyle\left\vert E \left\{\prod_{i=1}^{n} Z_i\right\}- E\{Z_1\}E \left\{\prod_{i=2}^{n} Z_i\right\} \right\vert + \left\vert  E\{Z_1\}E\left\{\prod_{i=2}^{n} Z_i\right\} -  \prod_{i=1}^{n} E \{Z_i\}\right\vert\\
&\leq&\displaystyle\left\vert E \left\{\prod_{i=1}^{n} Z_i\right\}- E\{Z_1\}E \left\{\prod_{i=2}^{n} Z_i \right\}  \right\vert + \left\vert E \left\{\prod_{i=2}^{n} Z_i\right\}- \prod_{i=2}^{n} E \{Z_i\} \right\vert.\\
\end{eqnarray*}
\end{flushright}

In the same way, we can bound the second term and finally we obtain

$$\left\vert E \left\{\prod_{i=1}^{n} Z_i\right\}- \prod_{i=1}^{n} E \left\{Z_i \right\}  \right\vert\leq\hspace{8cm}$$
\begin{flushleft}
$\displaystyle \leq\left\vert E \left\{\prod_{i=1}^{n} Z_i\right\}- E\{Z_1\}E \left\{\prod_{i=2}^{n} Z_i \right\}  \right\vert + \left\vert E \left\{\prod_{i=2}^{n} Z_i\right\}-E\{Z_2\}\;\prod_{i=3}^{n} E \{Z_i\}\right\vert +$
\end{flushleft}
\begin{flushright}
$\displaystyle\cdots+\left\vert E \left\{\prod_{i=n-1}^{n} Z_i\right\}- \prod_{n-1}^{n} E \{Z_i\}  \right\vert\hspace{2cm}$
\end{flushright}
\begin{flushleft}
$\displaystyle \leq\sum_{j=1}^{n-1}\left\vert E \left\{\prod_{i=j}^{n} Z_i\right\}-E\{Z_j\}\;E \left\{\prod_{i=j+1}^{n} Z_i\right\}\right\vert.$
\end{flushleft}
\end{dem'}

\thebibliography\\

\bibitem{Bernshtein}Bernshtein (1944). Extension of the central limit theorem of probability theory to sums of dependent random variables. {\em Uspehi Mat. Nauk} 10, 65-114 (in Russian).
\bibitem{Billingsley2}Billignsley, P. (1968). {\it Convergence of Probability Measures}. New York.  Wiley \& Sons.
\bibitem{Dobrushin}Dobrushin, P.L. (1968). The description of a random field by its conditional distribution. {\em Theory Probab. Appl.} 13, 201-229.
\bibitem{Doukhan}Doukhan (1995). {\em Mixing:Properties and Examples}.  Lectures Notes in Statistics 85, Springer Verlag.
\bibitem{Doukhan}Doukhan \& Louichi (1996). Weak dependence and moment inequalities. {\em Universit\'e de Paris-Sud, Pr\'epublication 97.08}.
\bibitem{Feller}Feller, W. (1978). {\it Introducci\'on a la Teor\'{\i}a de Probabilidades y sus Aplicaciones.} Vol.II. M\'exico. Limusa.
\bibitem{Perera1}Perera G. (1994)a). Estad\'{\i}stica Espacial y Teoremas  Centrales del L\'{\i}mite. Tesis doctoral, Centro de Matem\'atica, Universidad de la Rep\'ublica, Uruguay.
\bibitem{Perera2}Perera G. (1994)b).  Spatial Statistics, central limit theorems for mixing random fields and the geometry of $\mathbb{Z}^d$. {\em C.R. Acad. Sci. Paris} t.319, S\'erie I, 1083-1088.
\bibitem{Perera3}Perera G. (1997). Geometry of $\mathbb{Z}^d$ and the Central Limit Theorem for weakly dependent random fields. {\em Journal of Theoretical Probability}, Vol. 10, No. 3, 581-603.
\bibitem{Perera4}Perera G. (2000).{\it Random Fields on $\mathbb{Z}^d$, Limit Theorems and Irregular Sets}. Lecture Notes in Statistics Nro. 159:57-78. Centre de Recherches Math\'ematiques. Springer.

\end{document}